# Strong Off-Resonant Enhancement of Spontaneous Emission in Metal-Dielectric-Metal Plasmon Waveguide Structures

Y.C. Jun, R.D. Kekapture, J.S. White, and M.L. Brongersma

*Geballe Laboratory for Advanced Materials, 476 Lomita Mall, Stanford, California 94305*

We theoretically investigate the spontaneous emission process of an optical, dipolar emitter in metal-dielectric-metal slab and slot waveguide structures. We find that both structures exhibit strong off-resonant emission enhancements due to the tight confinement of modes between two metallic plates. The large enhancement of surface plasmon-polariton excitation enables dipole emission to be preferentially coupled into plasmon waveguide modes. These structures find applications in creating nanoscale local light sources or in generating guided single plasmons in integrated optical circuits.

Understanding the interactions between single emitters and surrounding electromagnetic fields has been of great importance for both fundamental studies and device applications. Coherent interactions between atoms and fields provide an ideal test bed for studying fundamental aspects of quantum mechanics [1,2], while the large enhancements of spontaneous emission (SE) in resonant cavities have enabled more efficient light emitting devices, drastic reductions in the threshold of lasers, and high efficiency single photon sources [3,4]. In addition to the many studies of sophisticated dielectric cavity and waveguide systems, there have been a number of investigations on SE enhancement near simple metallic structures such as metallic films [5,6] or nanoparticles [7,8]. These structures exploit the large density of states at the surface plasmon resonance frequency to achieve large emission enhancements. However, these resonant enhancements are limited by the narrow bandwidth and large metal losses around the surface plasmon resonance frequency.

In this paper, we study the spontaneous emission of a single emitter in metal-dielectric-metal (MDM) *slab* (2D) and *slot* (1D) waveguide structures (Fig. 1) that support bound surface plasmon-polariton (SPP) modes over extremely broad wavelength regions [9-11]. Both structures are shown to exhibit strong off-resonant emission enhancement due to the tight confinement of modes between two metallic plates, resulting in preferential coupling of the dipole emission into SPP waveguide modes. In off-resonant regions, propagation lengths of SPP modes become longer. Additionally, MDM structures can be reproducibly fabricated and integrated with other chip-scale components. Such structures can thus be useful for building efficient on-chip light sources for integrated optics. Coupling the emission of single emitters to a well-defined waveguide mode may be also useful for generating and guiding single plasmons (or single photons) in an optical circuit.

A MDM slab (Fig. 1(a)) supports one fundamental TM mode (also called the gap SPP mode) for sufficiently small gap sizes (~ order of tens of nanometers) [9,10], making it close to an ideal 2-dimensional waveguide. Because the electric field inside the gap is primarily directed normal to the metal surfaces, we expect that only a dipole oscillating normal to the metal surfaces will strongly couple to this mode. The SE enhancement factor $F_P$ of such a dipole in the gap can be obtained by considering the work done on the dipole by its own reflected field [12, 13]:

$$F_P = \frac{\gamma_\perp}{\gamma_0} = \frac{3}{2} \operatorname{Im} \int_0^\infty \frac{(1 - r_{12}^P e^{i\phi_{t2}})(1 - r_{13}^P e^{i\phi_{t3}})}{(1 - r_{12}^P r_{13}^P e^{i\phi_{t2}} e^{i\phi_{t3}})} \frac{u_\parallel^3}{\ell_1} du_\parallel \quad (1)$$

where the indexes 1,2,3 indicate the dielectric and two semi-infinite metal regions. The decay rate of a normal dipole ($\gamma_\perp$) is normalized to that in a uniform background dielectric medium

($\gamma_0$). Eq. (1) is derived from a plane wave decomposition of the emitted waves and $u_\parallel = k_\parallel / k_1$, $\ell_1 = -i(1-u_\parallel^2)^{1/2}$ correspond to their normalized in-plane and out-of-plane wavevectors. The Fresnel reflection coefficients ($r_{12}^p, r_{13}^p$) are multiplied by the corresponding round-trip phase changes ($e^{i\phi_{12}}, e^{i\phi_{13}}$). The frequency-dependent dielectric constant of metal is obtained from the literature [15-17] and we assume internal quantum efficiency $\eta_0 = 1$ for the emitter.

Figure 2 shows the calculated SE enhancement factor as a function of the free-space wavelength for a dipole in the center of the gap, where the fields are exactly perpendicular to the metal surfaces. In addition to a resonant enhancement peak around 400 nm (which is close to surface plasmon resonance wavelength of silver), there is strong off-resonant enhancement which increases linearly with wavelength (Fig. 2(a)). For comparison, the SE enhancement factor for an emitter spaced by the same distance from a single metal surface is shown as well, which clearly lacks such off-resonant enhancement. We also observe that as the gap size decreases, both resonant and off-resonant enhancements increase rapidly (Fig. 2(b)). To identify what causes these enhancements, a plot of the decay rate density (which is the integrand of Eq. (1)) as a function of normalized in-plane wavevector is shown (inset of Fig. 2(c)). We find that decay rate density spectra for MDM slabs are dominated by single peaks with wavelength-dependent peak locations. By plotting the positions of these peaks along with the gap SPP mode dispersion curve, it can be seen that these peaks lie exactly on the curve (Fig. 2(c)). This shows that the SE of a dipole inside the MDM slab is dominated by gap SPP excitation. The decay rate density also has contributions from broad, large wavevector components, so-called 'lossy surface waves' (LSW) that originate from intrinsic metal losses [13,14]. By integrating the relevant wavevector regions, we can estimate the fraction of energy coupled into the gap SPP mode among three different decay channels: gap SPP, LSW, and the conventional TM waveguide mode. Fig. 2(d) shows that gap SPP excitation is the dominant decay channel in a MDM slab for a wide range of metal-emitter distances at the considered off-resonant wavelength ($\lambda_0 = 800$ nm). We also note that when the metal-emitter distance becomes larger than ~ 250 nm, the fraction of gap SPP excitation decreases due to higher-order TM waveguide modes (which are similar to conventional waveguide modes). This behavior is in stark contrast to that observed for an emitter near a single metal surface, which exhibits relatively stronger coupling to LSW and free space modes.

To understand the physical origin of the behaviors observed in Fig. 2, we derive a simple analytical formula for SE enhancement for a dipole inside the gap of a MDM slab. From Fermi's golden rule, the SE rate of an emitter is given by $\gamma(x_e, \omega) = 2\pi |g(x_e, \omega)|^2 D_{2d}(\omega)$ where $|g|$ is the coupling strength between the dipole and the electromagnetic field at the emitter position $x_e$, and $D_{2d}$ is the density of states. Assuming the dipole $d_0$ is oscillating normal to the metal surfaces, the coupling strength is given by $|g(x_e, \omega)|^2 = |\vec{d}_0 \cdot \alpha \vec{E}(x_e, \omega)/\hbar|^2 = \omega |\vec{d}_0|^2 /(2\hbar \varepsilon \varepsilon_0 V_{eff})$, where $|\alpha|^2 = \hbar \omega / \int_{-\infty}^{\infty} \left\{ \varepsilon_0 \frac{d(\varepsilon \omega)}{d\omega} |E|^2 + \mu_0 |H|^2 \right\} dx$ is a normalization factor. We define the effective mode volume as $V_{eff} = L_{eff} \ell^2$, where $\ell$ is an arbitrary, in-plane quantization length and $L_{eff}$ is the effective mode length measured across the gap: $L_{eff}(x_e, \omega) = \frac{1}{2} \int_{-\infty}^{\infty} \left\{ \varepsilon_0 \frac{d(\varepsilon \omega)}{d\omega} |E(x,\omega)|^2 + \mu_0 |H(x,\omega)|^2 \right\} dx / (\varepsilon \varepsilon_0 |E(x_e, \omega)|^2)$. The density of states can be obtained by counting modes in a 2D space, and is given by $D_{2d}(\omega) = \ell^2 \omega / (2\pi v_p(\omega) v_g(\omega))$ where $v_p$ and $v_g$ are the phase and group velocity, respectively. By normalizing the SE rate with that in a uniform background medium ($\gamma_0(\omega) = \omega^3 \sqrt{\varepsilon} |d_0|^2 /(3\hbar \pi \varepsilon_0 c^3)$), we arrive at the expression for the SE enhancement factor due to the gap SPP excitation:

$$F_{SP} = \frac{\gamma_{sp}}{\gamma_0} = \frac{3}{4} \cdot \frac{c/n}{v_p} \cdot \frac{c/n}{v_g} \cdot \frac{\lambda_0/n}{L_{eff}} \qquad (2)$$

Figure 3 shows the calculated enhancement factor $F_{SP}$ as a function of wavelength. We find that Fermi's golden rule estimation retrieves the result of the analytical solutions well. Near the surface plasmon resonance, $v_p$ and $v_g$ rapidly decrease and thus $F_{SP}$ exhibits a peak. In the off-resonant region, the velocity reductions are small, but the normalized mode length $L_{eff}/(\lambda_0/n)$ decreases steadily with wavelengths due to the tight confinement of modes between the two metal plates, giving rise to substantial off-resonant enhancements. This is in contrast to a single metal surface which confines the mode tightly only near the surface plasmon resonance frequency. As the gap size is decreased, $v_p$, $v_g$, and $L_{eff}$ all decrease, thus giving larger enhancements. Finally, we find that this strong gap SPP excitation in MDM slab structures can enable it to be the main decay channel even down to very small metal-emitter distances, as shown in Fig. 2(d). Eq. (2) also explains the difference between real metal and ideal perfect electrical conductor (PEC). The plasmonic response of real metal reduces $v_p$ and $v_g$, which results in larger enhancement than that of PEC MDM (dotted line in Fig. 2(a)).

Strong SE enhancement can also be achieved with MDM slot structures which are more compatible with large scale integration. Recently, it was shown that slot structures (with critical dimensions on the order of tens of nanometers) support broadband and highly confined plasmonic modes – even into the long IR regime [11,18]. In order to quantify the attainable enhancement, we first obtain the eigenmodes of slot waveguides with COMSOL finite element simulations (Fig. 1(b)). We find that the mode is tightly confined around the slot and the electric field inside the gap is again primarily directed normal to the metal slot surface. Therefore, we expect that this slot waveguide structure can also support off-resonant enhancement for a dipole oriented normal to the metal surfaces. We derive a similar analytical formula for the SE enhancement factor in slot waveguides. The coupling strength is given by the same expression $|g|^2 = \omega|d_0|^2/(2\hbar\varepsilon\varepsilon_0 V_{eff})$, but now we define the effective mode volume as $V_{eff} = A_{eff} \cdot \ell$ where the effective mode area is $A_{eff}(\vec{r}_o,\omega) = \frac{1}{2}\iint\left\{\varepsilon_0\frac{d(\varepsilon\omega)}{d\omega}|E(\vec{r},\omega)|^2 + \mu_0|H(\vec{r},\omega)|^2\right\}d\vec{r}/(\varepsilon\varepsilon_0|E(\vec{r}_o,\omega)|^2)$. We assume a 1D density of states $D_{1d}(\omega) = \ell/(\pi v_g(\omega))$. From Fermi's golden rule, we find the emission enhancement due to the slot mode excitation to be:

$$F_{SP} = \frac{\gamma_{sp}}{\gamma_0} = \frac{3}{4\pi}\cdot\frac{c/n}{v_g}\cdot\frac{(\lambda_0/n)^2}{A_{eff}} \qquad (3)$$

From COMSOL finite element simulations, we solved for the group velocity and mode area of slot eigenmodes for three different metal film thicknesses (Fig. 4(b)) with a fixed gap size. The calculated enhancement factors (Fig. 4(a)) show that the simple slot structures exhibit strong off-resonant enhancements due to a large mode area reduction. Additionally, we see that as the film gets thinner, the enhancement increases due to both group velocity and mode area reductions. This large enhancement of slot mode excitation brings up the possibility that slot mode excitation can be a dominant emission channel in a MDM slot over other decay channels such as free-space radiation, loss to LSW, and SPP excitation (propagating along separate metal plates).

To verify our estimation, full-field 3D finite-difference time domain (FDTD) calculations were used, which include all decay pathways. The enhancement factor can be obtained by calculating the power dissipated by a dipole in the center of a MDM slot and in free-space: $F_P = \gamma/\gamma_0 = \langle P\rangle/\langle P_0\rangle$, where $\langle P(t)\rangle = \frac{\omega}{2}\text{Im}\{\vec{d}_0\cdot\vec{E}(r_o)\}$ [13]. First, as a validity check, the result of 3D FDTD calculations are compared to the analytic case of an infinite thickness slot (i.e. MDM slab) (Fig. 5(a)). The FDTD calculation retrieves the analytic results in both resonant and off-resonant regions, although it is slightly overestimated, likely due to the finite space discretization of 2 nm used in the simulations. 3D FDTD calculations were performed for MDM slots of three different metal film thicknesses (Fig. 5(b)). From the simulated profile of the Poynting vector, it can be clearly seen that a bound mode is launched into the slot direction (inset of Fig. 5(b)). Strong emission enhancements are indeed observed at off-resonant wavelengths. Furthermore, the film thickness dependence of SE enhancement agrees with the previous estimation. As the enhancement values in Fig. 4(a) (due to the slot mode excitation) are only slightly smaller than those in Fig. 5(b) (including all decay pathways), we can infer that the slot mode excitation is the main decay channel in slot waveguides over a broad range of wavelengths. The fraction of energy coupled to the slot mode can be estimated more directly through 3D FDTD flux calculations. By computing the flux into the slot direction and normalizing it to the total power flux, the coupling efficiency to the slot mode can be determined ($\beta_{SLOT} = \gamma_{SLOT} / \gamma_{TOT} = \langle P_{SLOT} \rangle / \langle P_{TOT} \rangle$) [19]. For L = 40 nm and t = 50 nm, we find the slot mode coupling efficiencies ($\beta_{SLOT}$) are ~ 80-90% at the considered off-resonant wavelengths.

In the off-resonant regime, the electromagnetic field penetrates less into the metal and the propagation lengths of the plasmon modes in both MDM slab and slot structures rapidly increase with wavelength [9,10]. Consequently, the large off-resonant enhancement is a desirable feature for optical device applications involving SPP modes. Moreover, in the off-resonant regime, MDM structures confine the electromagnetic mode mainly in the dielectric region between the metal plates, enabling efficient conversion of the SPP mode to a conventional dielectric waveguide mode through proper coupler structures [20,21]. By combining these out-coupling structures with MDM waveguides, MDM structures can be used for on-chip, local light sources in highly integrated optical systems.

Coupling the emission of single emitters to a slot waveguide enables efficient ways to generate and guide single plasmons on an integrated optical circuit, which can be also converted to single photons in free-space or in a dielectric waveguide [22,23]. MDM slots can be fabricated reproducibly with standard nanofabrication techniques (such as focused ion beam) and can be easily integrated with other components. The broadband nature of these structures makes spectral tuning unnecessary, thus removing stringent fabrication requirements to get reproducible emission enhancements and making it robust against fabrication imperfections. Additionally, the two separate metal plates of MDM structures can be used for electrical contacts and may enable tunable single photon emission or switching by applying electric fields to an emitter in the gap. Finally, MDM slot waveguides are also expected to be useful for studying the recently proposed nonlinear interactions of single photons via two-level [24] or three-level [25] atomic systems in a 1D waveguide.

In conclusion, we have shown that simple MDM waveguide structures can support strong, off-resonant emission enhancements over broad wavelength regimes. The resulting efficient coupling of emission to the plasmon mode makes MDM structures promising for both fundamental light-matter interaction studies and device applications.

The authors wish to thank Ed Bernard, Shanhui Fan, and J.T.Shen for helpful discussions. This work was funded by the DOE (grant: F49550-04-10437). Young Chul acknowledges the support of Samsung Scholarship.

* Corresponding author: ycjun@stanford.edu, brongersma@stanford.edu

[19] The flux into the slot direction decreases as away from the source, due to the metal loss. To compensate for this, the flux was measured as a function of distance to the source and fitted to an exponential function ( $e^{-x/L_p}$ ) to get the undamped flux value, where $L_p$ is the propagation length and $x$ is the distance to the source.

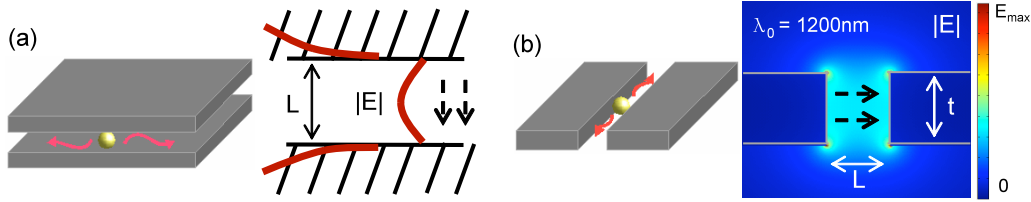

**Fig. 1** (color online). Schematic of metal-dielectric-metal (a) slab and (b) slot waveguides and electric field profile (|E|) of their fundamental modes. Dotted arrows indicate electric field directions inside the gaps. In (a), two semi-infinite metal plates are separated by a dielectric region of width L. In (b), two thin metal plates of thickness t are separated by a distance L and embedded in a uniform dielectric. The metal is silver and the dielectric has $\varepsilon_{dielectric} = 2.25$.

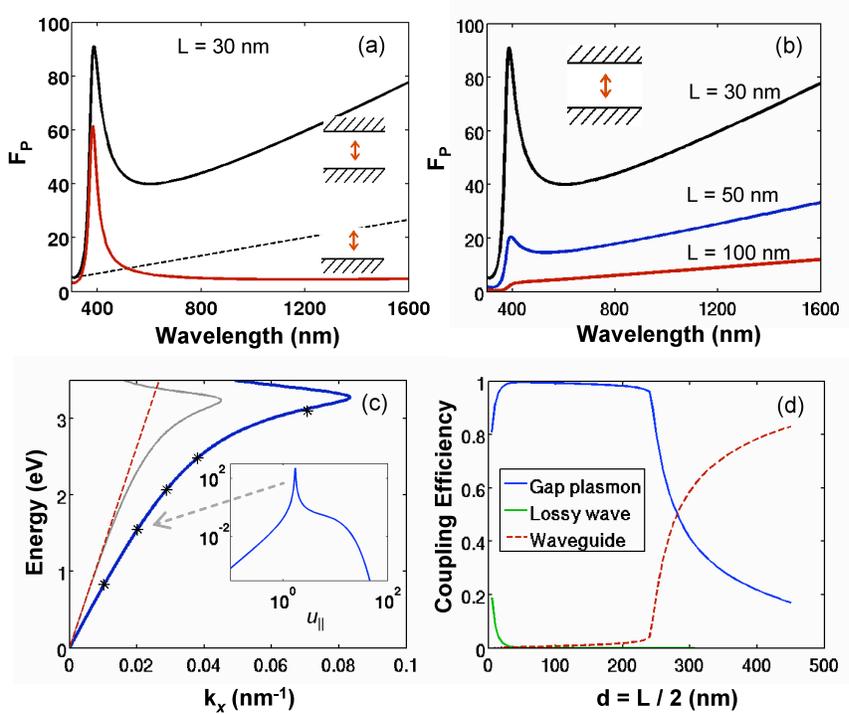

**Fig. 2** (color online). Spontaneous emission enhancement factor as a function of free-space wavelength ($\lambda_0$) (a) for a MDM structure vs. a single metal-dielectric interface, and (b) for a MDM structure with different gap sizes. (c) Dispersion relation of the gap plasmon mode for L = 30nm (blue), and the dispersion of a single silver surface (gray) and light line (dotted red). Inset:

decay rate density as a function of normalized in-plane wavevector. (d) Fraction of dissipated energy to each decay path as a function of metal-emitter distance d at $\lambda_0$ = 800 nm. In all cases, the emitter is in the middle of the gap. In (a), the enhancement factor for a PEC MDM is drawn with the dotted line.

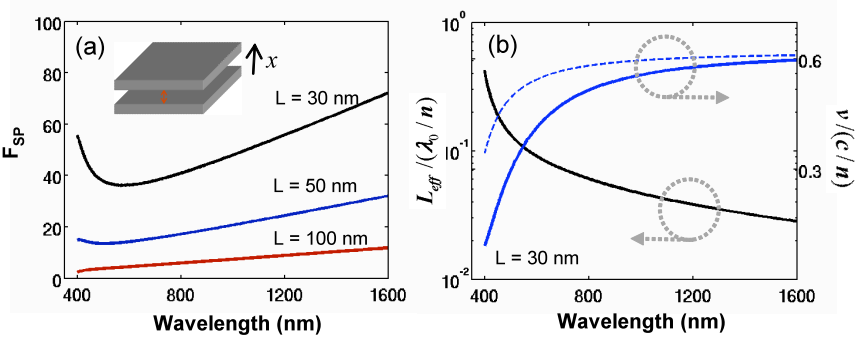

**Fig. 3** (color online). (a) Emission enhancement factor for a MDM slab (due to the gap SPP excitation) as a function of $\lambda_0$. (b) Normalized group velocity (blue), phase velocity (dotted blue), and mode length (black) for L = 30 nm.

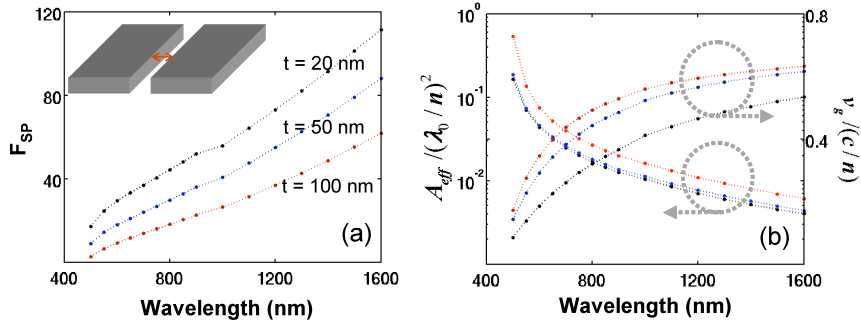

**Fig. 4** (color online). (a) Emission enhancement factor for a MDM slot (due to the slot mode excitation) as a function of $\lambda_0$ for different metal thicknesses, and L = 40 nm. (b) Normalized group velocity and mode area (black: t = 20 nm, blue: t = 50 nm, red: t = 100 nm).

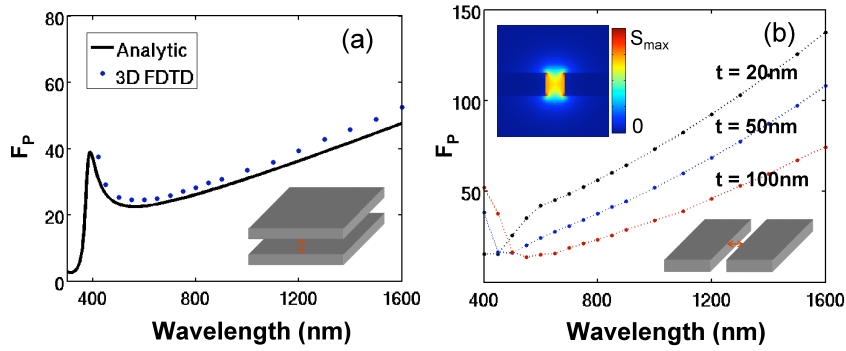

**Fig. 5** (color online). (a) Spontaneous emission enhancement factor for a MDM slab. 3D FDTD results are compared to the analytical solution. (b) Spontaneous emission enhancement factor for a MDM slot (calculated from 3D FDTD simulation for three different metal thicknesses, and L = 40nm). Inset: Poynting vector profile ($|\langle S \rangle|$) calculated 600 nm away from the source dipole (for t = 50nm and $\lambda_0$ = 1200 nm).